\documentclass[fleqn,10pt]{article}
\usepackage[dvips]{graphicx}
\begin{document}
\title{
\bf
Photoproduction of $\eta$ mesons on protons in the resonance region:
The background problem and the third $S_{11}$ resonance.
}
\author{V.A. Tryasuchev}
\date{\today}
\maketitle

\begin{abstract}
We have constructed an isobar model for the $\eta$-photoproduction on the proton in
the energy region up to the photon lab energy $K_0 = 3$ GeV. The data base involved
into the fitting procedure includes precise results for the cross section and for
the $T$-asymmetry of the process $\gamma p\to\eta p$ near threshold obtained at MAMI
and ELSA as well as recent results for the $\Sigma$-asymmetry and for the angular
distribution measured at higher energies in Grenoble and also more recent measurements
performed at JLab for the photon energies up to 2 GeV. The model includes twelve
nucleon resonances: $S_{11}(1535)$, $S_{11}(1650)$, $S_{11}(1825)$, $P_{11}(1440)$,
$P_{13}(1720)$, $D_{13}(1520)$, $D_{15}(1675)$, $F_{15}(1680)$, $F_{17}(1990)$,
$G_{17}(2190)$, $G_{19}(2250)$, $H_{19}(2220)$, and the background consisting of the
nucleon pole term and the vector meson exchange in the $t$-channel. To explain the
observed energy dependence of the integrated cross section, two $s$-wave resonances,
$S_{11}(1650)$ and $S_{11}(1825)$, have to be taken into account along with the
dominating $S_{11}(1535)$. The integrated cross section as well as the angular
distribution and $\Sigma$ asymmetry predicted by the model are in good agreement
with the data. Above the photon energy $K_0 = 2$ GeV, the calculated cross section
exhibits an appreciable dependence on the $\rho$- and $\omega$-meson contribution,
whose coupling with nucleons is not well defined. Several versions of extending the
model to higher energies are considered.
\end{abstract}

\section{Introduction}

Information on the electromagnetic properties of the nucleon resonances are
necessary for constructing the dynamical quark model of
baryons. Photoproduction of pions on nucleons and nuclei remains a
primary source of such information. On the other hand, because of a variety of
$N^*$ and $\Delta$ resonances excited in the pion production, it is rather
difficult to extract unambiguously their properties from the measured
observables. Contrary to the pions, the $\eta$-photoproduction
\begin{equation}\label{10}
\gamma+p\to\eta+p
\end{equation}
involves only the $N^*$ resonances with isospin $I=\frac{1}{2}$, which
make it possible to separate their contribution from the nearby
$\Delta$ resonances ($I=\frac{3}{2}$), appearing in the reactions with pions.
It should also be noted that in spite of the common resemblance of  process
(\ref{10}) to the pion photoproduction
on a nucleon, there is a principle difference: Whereas the $\pi$-meson
consists of the $u$- and $d$-quarks, the $\eta$-meson includes also the
strange quark-antiquark pair.

Apart from obtaining the information on the electromagnetic properties of the
nucleon resonances, sophisticated models for process (\ref{10}) are also
needed to study the $\eta$-photoproduction on nuclei, which in turn is
necessary for an investigation of the $\eta$-nuclear interaction. The modern
$\eta$ photoproduction analysis can also be useful for estimating various
background reactions in experiments with $\eta$ mesons.

\section{Description of a model}

In the center-of-mass system, the amplitude $F$ of photoproduction of a
pseudoscalar meson on a nucleon reads \cite{CGLN}
\begin{equation}\label{20}
F=iF_1\vec{\sigma}\cdot\vec{\varepsilon}
+ F_2\vec{\sigma}\cdot\hat{q}(\hat{k}\times\vec{\varepsilon}\,)
+iF_3\vec{\sigma}\cdot\hat{k}\,\hat{q}\cdot\vec{\varepsilon}
+iF_4\vec{\sigma}\cdot\hat{q}\,\hat{q}\cdot\vec{\varepsilon}\,,
\end{equation}
where $\vec{\sigma}$ are Pauli matrices, representing the nucleon spin
operator; $\hat{k}=\vec{k}/k$ and $\hat{q}=\vec{q}/q$ with $\vec{k}$ and
$\vec{q}$ being the photon and $\eta$ c.m.\ momenta, and the unit vector
$\vec{\varepsilon}$ determines the photon polarization. The four values $F_i$
are functions of the total energy
$W$ and the meson c.m.\ angle $\theta$. In the angular momentum
representation,
$F$'s are expressed in terms of electric $E_{l\pm}$ and magnetic $M_{l\pm}$
multipole amplitudes \cite{CGLN,BlWei52}
\begin{eqnarray}
F_1&=&\sum\limits_{l=0}^\infty\left[lM_{l^+}+E_{l^+}\right]\,P'_{l+1}(x)
+\sum\limits_{l=0}^\infty\left[(l+1)M_{l^-}+E_{l^-}\right]\,P'_{l-1}(x)\,,
\nonumber\\
F_2&=&\sum\limits_{l=1}^\infty\left[(l+1)M_{l^+}+lM_{l^-}\right]\,P'_l(x)\,,\\
F_3&=&\sum\limits_{l=1}^\infty\left[-M_{l^+}+E_{l^+}\right]\,P''_{l+1}(x)
+\sum\limits_{l=3}^\infty\left[M_{l^-}+E_{l^-}\right]\,P''_{l-1}(x)\,,
\nonumber\\
F_4&=&\sum\limits_{l=1}^\infty\left[M_{l^+}-E_{l^+}-M_{l^-}
-E_{l^-}\right]\,P''_l(x),\nonumber
\end{eqnarray}
where $P_l(x)=P_l(\cos\theta)$ are Legendre polynomials. The observables of 
process (\ref{10}) can be expressed in terms of the amplitudes $F_i$ as follows.
Differential cross section:
\begin{eqnarray}
\frac{d\sigma}{d\Omega}&=&\frac{q}{k}\left\{|F_1|^2+|F_2|^2-
2{\cal R}e(F_1F_2^*)\cos\theta\right. \nonumber\\
&+&\left.\frac{1}{2}\left[|F_3|^2+|F_4|^2+
2{\cal R}e(F_1F_4^*)+2{\cal R}e(F_2F_3^*)+2{\cal R}e(F_3F_4^*)\cos\theta\right]
\sin^2\theta\right\}\,.
\end{eqnarray}
$\Sigma$-asymmetry due to the linear photon beam polarization parallel and
perpendicular to the reaction plane:
\begin{equation}\label{32}
\Sigma\frac{d\sigma}{d\Omega}=-\frac{q}{k}\frac{\sin^2\theta}{2}
\left[|F_3|^2+|F_4|^2+
2{\cal R}e(F_2F_3^*+F_1F_4^*+F_3F_4^*\cos\theta)\right]\,.
\end{equation}
$T$-asymmetry due to the initial nucleon polarization  parallel and
perpendicular to the reaction plane:
\begin{equation}\label{40}
T\frac{d\sigma}{d\Omega}=\frac{q}{k}\sin\theta\,{\cal I}m
\left[F_1F_3^*-F_2F_4^*+(F_1F_4^*-F_2F_3^*)\cos\theta
-F_3F_4^*\sin^2\theta\right]\,.
\end{equation}
$P$-asymmetry due to the final nucleon polarization:
\begin{equation}\label{50}
P\frac{d\sigma}{d\Omega}=\frac{q}{k}\sin\theta\,{\cal I}m
\left[F_2F_4^*-2F_1F_2^*-F_1F_3^*+(F_2F_3^*-F_1F_4^*)\cos\theta
+F_3F_4^*\sin^2\theta\right]\,.
\end{equation}
Nowadays there are rigorous models for process (\ref{10}) which
describe very accurately the cross section in the region near threshold
\cite{BeTa91,BeMu91,TiBeKa94,BeMu95,FiAr97}.
At higher photon energy, due to relatively large $\eta$ mass,
the higher resonances corresponding to the states with larger angular momentum
are expected to be intensively
involved into the production mechanism. Such states can
simply be included into the model by using the Breit-Wigner ansatz for the
$s$-channel part of the corresponding partial amplitudes. Therefore, the
resonance multipoles were taken in the form
\cite{HiDe73}
\begin{equation}\label{60}
E_{l^\pm}=\frac{e^{i\Phi_r}\left(\Gamma^E_{\gamma p}
\Gamma_{\eta p}\right)^{1/2}}{2\left[ kq\,j(j+1)\right]^{1/2}\left(
W_r-W-\frac{i}{2}\Gamma\right)}
\end{equation}
and the analogical expression for the magnetic amplitudes $M_{l\pm}$. Here $l$
stands for the meson angular momentum, and
$j=l\pm 1$ for $E_{l\pm}$ and $j=l$ for
$M_{l\pm}$ respectively. The total $\eta N$ angular momentum is
$J=l\pm\frac{1}{2}$. The energy $W_r$ is the total energy of the $\eta N$
system at the resonance position.
Energy dependence of the total width $\Gamma$ was chosen
according to the prescription given in Ref.~\cite{HiDe73}
\begin{equation}\label{80}
\Gamma=\sum\limits_\alpha\frac{p\nu_l(pR)}{p_r\nu_l(p_r R)}
\Gamma_{\alpha r}\,,
\end{equation}
where $\Gamma_{\alpha r}$ is the partial decay width in the channel $\alpha$,
and  the momenta $p$
and $p_r$ are the meson c.m.\ momenta corresponding to the energies $W$
and $W_r$, respectively. The function
$\nu_l(x)$ stands for penetration factors, associated
with the partial wave $l$. For $l\leq 3$, the needed
expressions for $\nu_l(x)$ are
given in \cite{BlWei52}. Since for $l>3$ these factors are rarely presented in
the literature we list here the expressions for $\nu_l(x)$
used in our model
\begin{eqnarray}\label{82}
\nu_0(x) & = &1\,,\nonumber \\
\nu_1(x) & = &\frac{x^2}{1+x^2}\,,\nonumber\\
\nu_2(x) & = &\frac{x^4}{9+3x^2+x^4}\,,\\
\nu_3(x) & = &\frac{x^6}{225+45x^2+6x^4+x^6}\,,\nonumber\\
\nu_4(x) & = &\frac{x^8}{11025+1575x^2+135x^4+10x^6+x^8}\,,\nonumber\\
\nu_5(x) & = &\frac{x^{10}}{893025+99225x^2+6300x^4+315x^6+15x^8+x^{10}}\,.
\nonumber
\end{eqnarray}
In Eq.\ (\ref{80}) the effective radius of
$\eta N$-interaction $R$, was taken to be equal to 1\,fm and was not varied in
the fitting procedure. For all resonances, but $S_{11}(1535)$, the sum in
(\ref{80}) was restricted to the pion channel. For the $S_{11}(1535)$ resonance
which quite intensively decays into the $\eta N$ channel the energy
dependent width was taken in the form
\begin{equation}\label{90}
\Gamma=\left(0.5\frac{q}{q_r}+0.4\frac{p}{p_r}+0.1\right)\Gamma_r\,,
\end{equation}
where $q$ and $p$ are c.m.\ momenta of $\eta$ and $\pi$ mesons, respectively,
and the corresponding values $\Gamma_r, q_r$, and $p_r$ are calculated at the
resonance position $W=W_r$.

The electromagnetic $\Gamma_{\gamma p}^{E,M}$ and strong $\Gamma_{\eta p}$
vertices of the resonance amplitudes (\ref{60}) were parametrized in
the form \cite{HiDe73}
\begin{equation}\label{100}
(\Gamma_{\gamma p}^{E,M}\Gamma_{\eta p})^{1/2}=\left\{\left[2kR\nu_n(kR)\right]\left[2qR\nu_l(qR)\right]\right\}^{1/2}\gamma^{E,M}
\end{equation}
where $n=l$, except for the multipole amplitudes $E_{l^-}$, for which
$n=l-2$. The values $\gamma^{E,M}$ were treated as free parameters together
with the main resonance characteristics $W_r$ and $\Gamma_r$.

Apart from the resonance terms, smooth background contributions
consisting of the nucleon pole terms in the $s$- and $u$-channel
and exchange of $\rho$- and $\omega$-mesons in the $t$-channel were
included into the model. The nucleon pole amplitudes read \cite{TiBeKa94}
\begin{eqnarray}\label{Born}
F_1&=&eg_{\eta NN}\frac{W+m}{8\pi W}\sqrt{\frac{E_f+m}{E_i+m}}
k\left[e_N+\mu_N\right]\left(\frac{1}{s-m^2}+\frac{1}{u-m^2}\right)
\,,\nonumber\\
F_2&=&-eg_{\eta NN}\frac{W-m}{8\pi W}
\sqrt{\frac{E_i+m}{E_f+m}}q\left[e_N+\mu_N\right]
\left(\frac{1}{s-m^2}+\frac{1}{u-m^2}\right)\,,\\
F_3&=&2eg_{\eta NN}\frac{W+m}{8\pi W}
\sqrt{\frac{E_f+m}{E_i+m}}qk\left[\frac{e_N}{W+m}+\frac{\mu_N}{2m}\right]
\frac{1}{u-m^2}\,,\nonumber\\
F_4&=&-2eg_{\eta NN}\frac{W-m}{8\pi W}
\sqrt{\frac{E_i+m}{E_f+m}}q^2\left[\frac{e_N}{W-m}-\frac{\mu_N}{2m}\right]
\frac{1}{u-m^2}\,,\nonumber
\end{eqnarray}
where $m$ is the nucleon mass; $g_{\eta NN}$ is the constant of the $\eta NN$
pseudoscalar coupling, and $e_N$ and $\mu_N$ stand for the charge
($e_p=1$, $e_n=0$) and the anomalous magnetic moment ($\mu_p=1.79,
\mu_n=-1.91$) of the nucleon; $e^2/4\pi$ is the fine-structure constant.

For the vector meson contribution one obtaines \cite{TiBeKa94}
\begin{eqnarray}\label{Vmeson}
F_1&=&e\lambda_V\frac{W+m}{8\pi W m_\eta}\sqrt{\frac{E_f+m}{E_i+m}}
k\left[\frac{G_{VNN}^t}{2m}-G_{VNN}^v\left(W-m+\frac{t-m_\eta^2}{2(W-m)}
\right)\right]\frac{1}{t-m_V^2}\,,\nonumber\\
F_2&=&-e\lambda_V\frac{W-m}{8\pi W m_\eta}\sqrt{\frac{E_i+m}{E_f+m}}
q\left[\frac{G_{VNN}^t}{2m}+G_{VNN}^v\left(W+m+\frac{t-m_\eta^2}{2(W+m)}
\right)\right]\frac{1}{t-m_V^2}\,,\\
F_3&=&-e\lambda_V\frac{W+m}{8\pi W m_\eta}\sqrt{\frac{E_f+m}{E_i+m}}
kq\left[\frac{G_{VNN}^t}{2m}(W-m)-G_{VNN}^v\right]\frac{1}{t-m_V^2}\,,
\nonumber\\
F_4&=&e\lambda_V\frac{W-m}{8\pi W m_\eta}\sqrt{\frac{E_i+m}{E_f+m}}
q^2\left[\frac{G_{VNN}^t}{2m}(W+m)+G_{VNN}^v\right]\frac{1}{t-m_V^2}\,.
\nonumber
\end{eqnarray}
Here $\lambda_V$ is the $\gamma \eta V$ coupling constant and the values
\begin{equation}
G_{VNN}^{v,t}=g_{VNN}^{v,t}G^V(t)
\end{equation}
are the constants determining vector and tensor $VNN$ coupling.

Our main task was to develop the model which could provide a good
description of the low energy data for $\gamma p\to \eta p$ and at the same
time has a "right"
asymptotic behavior at higher photon energies ($K_0 > 2$ GeV).

\section{Discussion of the results}

Varying the parameters of the resonances listed in Table \ref{tab1} within the
ranges recommended in Ref.\,\cite{PDG00} we obtain a good description of the data
for $\gamma p\to \eta p$. At the beginning also the relative phases $\Phi_r$ (see
Ref.\,\cite{TRS01}) were used in the fitting procedure, but after the eleven
resonances and the background terms were included into the model, we were able to
abandon these phase factors at all \cite{TRS03}. The resulting set of parameters
listed in Table~\ref{tab1} \cite{S11} was obtained by adjusting the model containing
all eleven resonances and the background to the low energy data ($K_0 < 1.1$ GeV)
presented in Ref.\,\cite{Kru95,Ren02,Ajak98,Bock98} as well as to the old
experimental results for $K_0 > 1.2$ GeV obtained in \cite{ABBHHM,Var98}.

\begin{figure}
\begin{center}
\includegraphics[width=10cm,keepaspectratio]{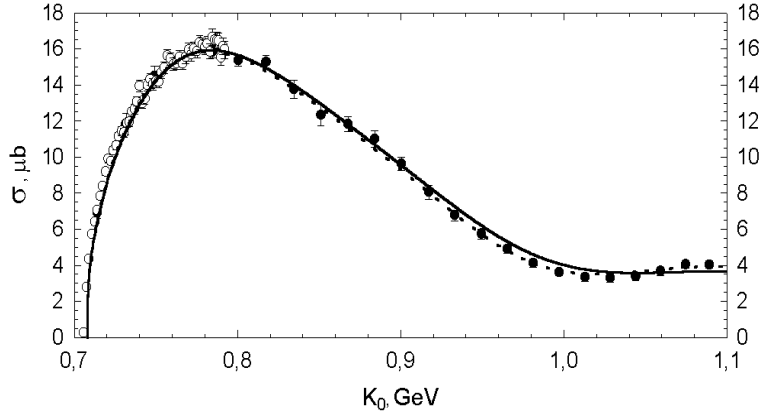}

\caption{ Integrated cross section for $\gamma p\to \eta p$ as a function of lab
photon energy $K_0$. The prediction of the present model is shown by the solid
curve. Doted curve is the result of the Ref.\,\protect\cite{TRS01}. The data are
from \protect\cite{Kru95} (open circles) and \protect\cite{Ren02} (filled circles).
} \label{fig1}
\end{center}
\end{figure}

As was repeatedly noted in the previous work, the dominant contribution to
the cross section of $\gamma p\to\eta p$ arises from the $S_{11}(1535)$
resonance, which is strongly coupled to the $\eta N$ channel. However
inclusion of this resonance alone does not reproduce energy dependence
exhibited by the measured total cross section. Only including
$S_{11}(1650)$, and damping by this means too a large contribution of
$S_{11}(1535)$ resonance
at the energies above 0.8 GeV, we are able to bring the theory
into agreement with the data which were
obtained in two experiments in different energy
regions (see Fig.\,\ref{fig1}).
Also the resonance $P_{13}(1720)$ provides within our model
an essential contribution to the
$\eta$ photoproduction amplitude.

Using the parameters of $S_{11}(1535)$ from Table~\ref{tab1} one obtains
\begin{equation}\label{110}
\Gamma_\gamma/\Gamma_r \approx 0.345\ \%\,,
\end{equation}
where $\Gamma_\gamma$ denotes the total radiative width of the resonance. This
result agrees with the maximal value of this ratio, recommended in
Ref.\,\cite{PDG00}. The same result can be presented in the form
\begin{equation}\label{120}
\Gamma_{\eta p}\Gamma_\gamma/\Gamma_r \approx 0.27\ \mbox{MeV}\,.
\end{equation}
Then, taking $\Gamma_{\eta p}/\Gamma_r = 0.50$, one obtains for the invariant
helicity amplitude $A_{1/2}=0.110$ GeV$^{-1/2}$. And conversely, taking
$A_{1/2}=0.090$ GeV$^{-1/2}$, as is recommended in \cite{PDG00},
one finds the ratio of the widths $\Gamma_{\eta p}/\Gamma_r = 0.74$. This result 
points to the
fact that the role of the $S_{11}(1535)$ resonance in the process
$\gamma N\to\eta N$ appears to be greater than the one prescribed by the PDG
analysis \cite{PDG00}. At the same
time, the value of the $S_{11}(1535)$ photoexcitation amplitude $A_{1/2}$ is
in good agreement with $A_{1/2}$ = 0.118 GeV$^{-1/2}$, obtained in
Ref.\,\cite{ChiNP02}.
As for the resonance $S_{11}(1650)$, if one takes $\Gamma_{\eta
 p}/\Gamma_r = 0.1$ then $\Gamma_\gamma/\Gamma_r \approx$ 0.38 $\%$ and
$A_{1/2} = 0.098$ GeV$^{-1/2}$. The latter value is about two times larger then the
upper limit, which can be found in Ref.\,\cite{PDG00}. For the resonance
$P_{13}(1720)$ the constants $\gamma^{E,M}$ obtained by fitting the data  
lead to the large absolute significance of the values $\xi_\lambda$ (see Table~\ref{tab1}). 
More detailed information about the electromagnetic vertices of $N^*$ resonances can be 
obtained from the results presented in Table~\ref{tab1}, if the corresponding ratios
 $\Gamma_{\eta p}/\Gamma_r$ are known.

Among the resonances with larger masses, only those marked in \cite{PDG00}
with four asterisks were included into the fitting procedure, namely
$F_{17}(1990)$,
$G_{17}(2190)$, $G_{19}(2250)$, and $H_{19}(2220)$. As the direct calculations
shows, insertion of these resonances influences mostly the $\Sigma$ asymmetry,
while the differential and total cross sections in the region $K_0 < 1.2$ GeV
are less affected. The most important contribution to $\gamma
p\to\eta p$ comes from the $G_{19}(2250)$ resonance.

\begin{figure}
\begin{center}
\includegraphics[width=18cm,keepaspectratio]{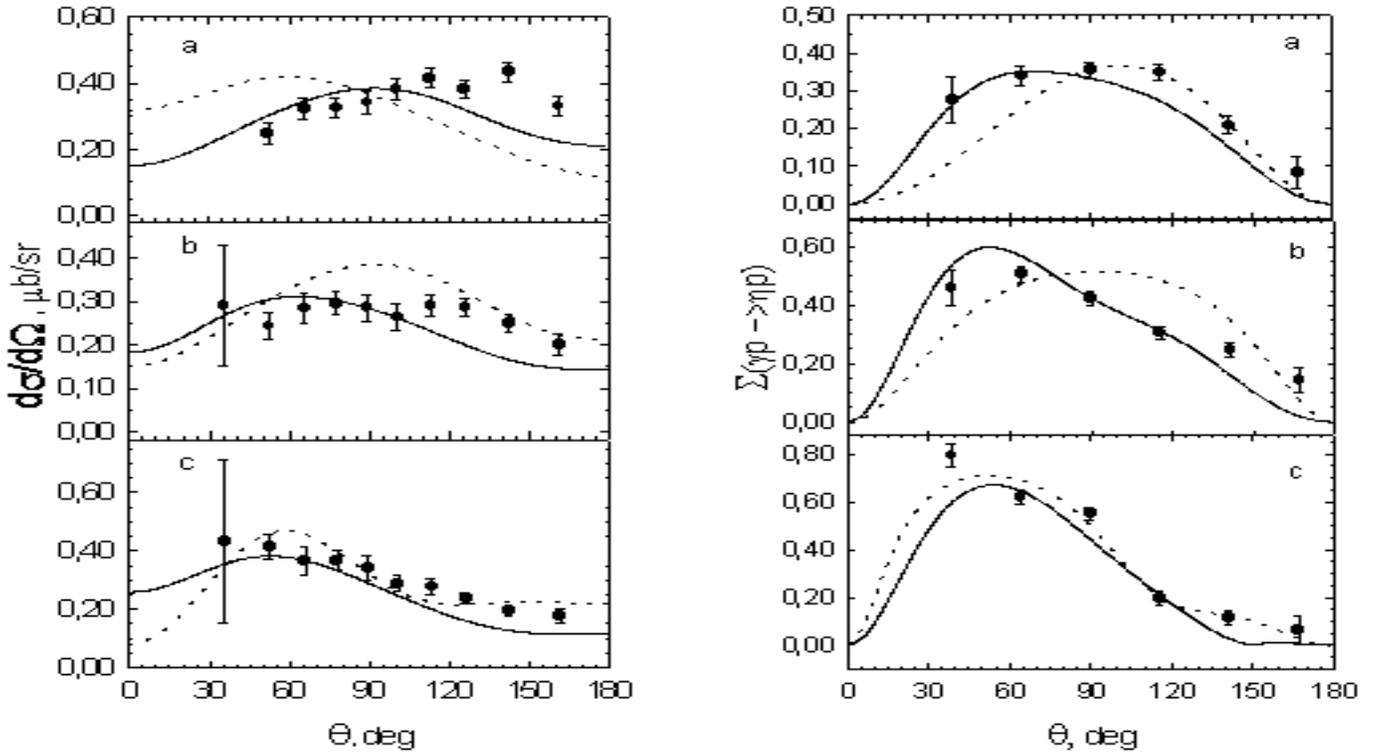}

\vspace{.5cm} \caption{Left panel: differential cross section $d\sigma/d\Omega$
versus the $\eta$ c.m. emission angle. Shown are the calculations at three lab
photon energies: $K_0
 = 981$ MeV (a), $K_0 = 1029$ MeV (b), and $K_0 = 1075$ MeV (c). The data are
 taken from \protect\cite{Ren02}. Solid curves are obtained using the model
 with eleven resonances with the parameters listed in Table~\protect\ref{tab1}.
The doted curves
 are the results of the model \protect\cite{TRS01}.
Right panel: The $\Sigma$ asymmetry calculated at $K_0=931$ MeV (a), $K_0=991$ MeV
(b), and $K_0=1056$ MeV (c). The data are from Ref.\,\protect\cite{Ajak98}. The
meaning of the curves as in the left panel. } \label{fig2}
\end{center}
\end{figure}

The calculated angular distributions $d\sigma/d\Omega$ are in excellent
agreement with the experimental results reported by MAMI (Mainz) and GRAAL
(Grenoble) up to the photon energy $K_0 \approx 950$ MeV. They were
discussed in our previous work \cite{TRS01} and are not presented here.
With increasing energy the theoretical distributions
$d\sigma/d\Omega$ exhibit pronounced decrease at forward angles (see also
Ref.\,\cite{ChiNP02}) which was not observed in the measurements \cite{Ren02}.
Some results around the energy $K_0 = 1$ GeV, which
were obtained in the present as well as in the previous work \cite{TRS01} are
compared with the data in Fig.\,\ref{fig2}.
It must be emphasized that the
model including eleven \cite{TRS03} resonances agrees with the experimental
results at these energies as good as the low energy model \cite{TRS01} with six
resonances. Furthermore, we would like to note that the maxima in $\Sigma
(\theta)$ for $K_0 > 950$ MeV appear at the same angles $(\theta\approx 50^0)$
as does the maxima of $d\sigma/d\Omega$ at the corresponding energies. This is
the reason why the experimental values of $\Sigma (\theta)$ increase strongly
at the energy around $K_0 = 1$ GeV (see Fig.\,\ref{fig2}).

\begin{figure}
\begin{center}
\includegraphics[width=10cm,keepaspectratio]{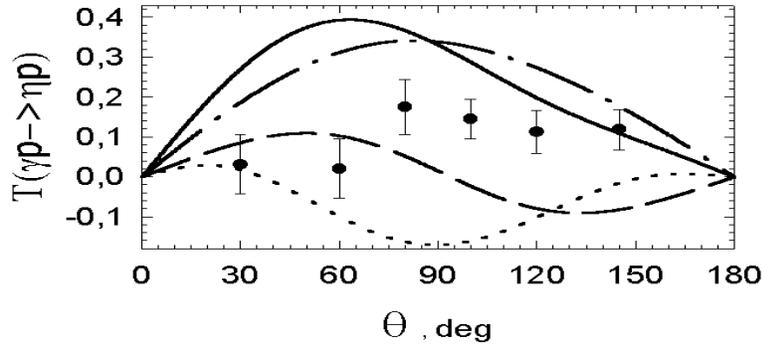}

\vspace{.5cm} \caption{ Target asymmetry for $\gamma p\to \eta p$ at $K_0 = 857$ MeV
calculated in the lab system. The theoretical results are obtained using the present
model with all resonances (solid curve), without $P_{13}(1720)$ (dashed curve), and
without $D_{15}(1675)$ (dash-dotted curve). The dotted curve presents the result of
Ref.~\protect\cite{TRS01}. The data are from \protect\cite{Bock98}. } \label{fig3}
\end{center}
\end{figure}

The target asymmetry $T(\theta)$ (\ref{40}) predicted by the model is
positive. Furthermore, it depends almost not at all on the born
(nucleon pole) term, as well as on the form factors in the $VNN$
vertices and agrees qualitatively with the Bonn results \cite{Bock98},
as is shown in Fig.\,\ref{fig3}.
In the same figure we also demonstrate the sensitivity of $T(\theta)$
to the contribution of the
$P_{13}(1720)$ and $D_{15}(1675)$ resonances. It must also be noted, that the
model of \cite{TRS01} (the result is shown by the doted curve) was in
disagreement with the experimental results. At the same time, we could not describe the 
experimental data of work \cite{Bock98} on the target asymmetry of process (1) for the 
photon lab energy below 750 MeV by using the resonances from table 1.

Finally, the polarization of the final protons $P(\theta)$ (\ref{50}) is in
strong disagreement with the experimental data obtained 30 years ago
\cite{Heu70}.

\section{Investigation of the background}

If the background contribution is not included, the integrated cross section
decreases rapidly to zero. Therefore, a pure resonance model for
$\gamma p\to\eta p$ seems to be incompatible with a real production dynamics.
As was noted above, the background contains the
nucleon pole term in $s$ and $u$ channels as well as $\rho$ and $\omega$
exchanges in $t$ channel. The nucleon pole term is defined by the $\eta NN$
coupling constant. Its contribution appears to be insignificant for all
reasonable values of $g_{\eta NN}$, therefore we take $g_{\eta NN}^2/4\pi =
0.4$ from Ref.\,\cite{TiBeKa94,FiAr97,ChiNP02}.

\begin{figure}
\begin{center}
\includegraphics[width=10cm,keepaspectratio]{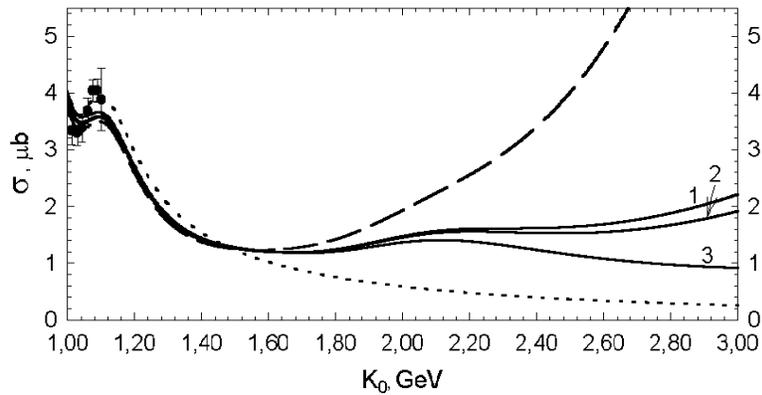}

\vspace{.5cm} \caption{ Integrated cross section for $\gamma p\to \eta p$ above the lab
photon energy $K_0 = 1$ GeV. The solid curves 1,2, and 3 are calculated with vector
meson exchange parameters from Refs.~\protect\cite{TiBeKa94}, \protect\cite{BeMu95}
and \protect\cite{ChiNP02}, respectively (see Table~\protect\ref{tab2}).The dashed
curve includes the vector meson exchange from Ref.~\protect\cite{TiBeKa94} but with
$G^V(t) = 1$.The dotted curve is the result of the model \protect\cite{TRS01}.
The data are taken from Ref.~\protect\cite{Ren02}. } \label{fig4}
\end{center}
\end{figure}

For pointlike $VNN$ vertices with $G^V(t) = 1$ the $\rho$ and $\omega$
contributions rise rapidly
with increasing photon energy so that for the photon energies above $K_0 =
1.5$ GeV the photoproduction amplitude is almost totally determined by the
vector meson exchange, and the cross section is anomalously enhanced
in the region close to $K_0 = 3$ GeV.
Taking into account a finite range of the $VNN$ coupling ($G^V(t)\to 0$
if $|t|\to \infty $) leads to visible damping of the cross section(see Fig.\,\ref{fig4}). 
However,its value depends strongly on the choice of
the $VNN$ form factors. In the present work we have considered two types of
form factors
\begin{equation}\label{130}
G^V(t)=\left(\frac{\Lambda_V^2-m_V^2}{\Lambda_V^2-t}\right)^n
\end{equation}
with $n=1$ (monopole type \cite{BeMu95}) and $n=2$ (dipole type
\cite{TiBeKa94,ChiNP02,ChiNT02}) with different cut off parameters $\Lambda_V$.
Whereas the constants $\lambda_V$ in Eq.\,(\ref{Vmeson}) are quite well known from
the radiation decays $V\to\gamma \eta$, (see, e.g., \cite{ChiNP02}), the hadronic
constants $g_{VNN}^v$ and $g_{VNN}^t$ are determined poorly for both $\rho$ and
$\omega$ mesons, and are usually treated as free parameters in phenomenological
models. Several sets of these constants, used in the previous work, are listed in
Table~\ref{tab2}. Appreciable difference between the total cross section obtained
with these sets appears only above the energy $K_0 = 1.9$ GeV (see
Fig.\,\ref{fig4}). It must be kept in mind that all but the latter sets given in
Table~\ref{tab2} were used for the description of $\gamma N\to\eta N$ observations
only in the low-energy region ($K_0 < 1.1$ GeV).

\section{Third $S_{11}$ resonance in $\gamma p\to\eta p$}

In the previous work \cite{TRS03} we have extended the model to higher energies
($K_0 > 1.2$ GeV) using the old database from \cite{ABBHHM,Var98},
since the experimental results obtained at JLab \cite{Dug02}
for $K_0 = 0.75 - 1.95$ GeV were published
later. The data \cite{Dug02} agree with those published in
\cite{Ren02} at $K_0 = 0.75 - 1.0$ GeV but above this energy region one notes a
visible deviation (Fig\,5). Our model with eleven resonances does not reproduce
the results of \cite{Dug02}. The same deviation
can also be observed by comparing the data with other analyzes, e.g
\cite{ChiNT02}.
In order to eliminate this disagreement we have introduced into the
model, apart from two $S_{11}$ resonances with masses 1535 and 1650 MeV,
the third $S_{11}$ resonance with the parameters given in Table~\ref{tab1}
(marked by asterisk).
Varying the ratio $\Gamma_{\eta p}/\Gamma_r$ in the region 0.5 to 0.01
one obtains for the invariant helicity amplitude of the third $S_{11}$
resonance the $A_{1/2}$ = 0.021 to 0.151 GeV$^{-1/2}$, appreciably smaller than
the corresponding values of other two $s$-wave resonances. The effect of
including the $S_{11}(1825)$ resonance into the model of Ref.\,\cite{TRS03}
is shown in Fig.\,\ref{fig6}. The contribution
vanishes totally at the ends of the interval $K_0 = 1.025 - 1.9$ GeV, as is
seen in Figs. 6a and 6d, which in turn guarantees that a quite
good description of the data achieved in the low-energy region within the
model with eleven resonances is not affected by insertion of the new
resonance.
There are appreciable differences between the differential cross sections
at forward and backward $\eta$ angles predicted by various models
\cite{ChiNP02,TRS03,Bai01,Sagh02}. Therefore
to discriminate between the models, measurements of the $\eta$
angular distribution at small ($\theta <
45^0$) and large angles ($\theta > 135^0$) became important. The main
distinctive property of the present model is,
that it takes explicitly into account the contribution of
higher resonances with large angular momenta, $F_{17}(1990)$, $G_{17}(2190)$,
$G_{19}(2250)$, $H_{19}(2220)$, which make the angular distribution of
$\gamma N\to\eta N$ at $K_0 > 1.5$ GeV strongly anisotropic (Fig.\ref{fig6}).
In this connection we take with some caution the estimation of the
total cross section made by the authors of Ref.\,\cite{Dug02}
who has used only the
values of $d\sigma/d\Omega$ in the interval $45^0 < \theta < 135^0$. As is
noted above, such estimation must depend strongly
on the model used for the extrapolation of the
cross section into the region of small and large $\eta$ angles.
The dependencies of the angular distributions of the $\eta$ mesons in process (1) for 
energy region $K_0 > 1.5$ GeV on contributions of higher-spin resonances are shown 
in Fig.\ref{fig7}. The differential cross section form depends significantly on
contributions of the resonances $F_{17}(1990)$, $G_{17}(2190)$, $G_{19}(2250)$.
The influence of the contribution of the resonance $H_{19}(2220)$ on it is minor.
In this energy region the calculations have shown that higher-spin resonances are
responsible for the large positive $\Sigma$ beam asymmetry. 

The third $s$-wave resonance discussed here is not heuristic. This resonance
was predicted by quark models and was also observed in other processes
\cite{Dug02,Bai01,Sagh02,CheKa02,GiSa01}.
Its parameters, needed to describe the data (see Table \ref{tab1}) are in
surprisingly good agreement with those, extracted from the experimental
results for the decay $J/\Psi\to p\overline{p}\eta$ \cite{Bai01}
($W_r = 1800\pm
40$ MeV, $\Gamma_r = 165^{+165}_{-85}$ MeV), but do not agree with the
properties of the third $S_{11}$ resonance predicted by the constituent
quark model with broken $SU(6)\otimes O(3)$ symmetry \cite{Sagh02}
($W_r = 1776$ MeV, $\Gamma_r = 268$ MeV).
While on the subject of the new $S_{11}$
resonances, we would like to note the work \cite{CheKa02}
where a unified analysis of the
data for the scattering and photoproduction of pions in a large energy
region was performed. The results of the cited work point to the existence of
the third and the forth $S_{11}$ resonances with masses $M_r=1846\pm 47$ MeV
and $M_r = 2113 \pm 70$ MeV respectively. However, the corresponding widths
can not be defined with such a precision and their values were estimated to be
more than 300 MeV. Although the authors of Ref.\,\cite{ChiNT02} do not introduce
additional $s$-wave resonance for the description of the $\gamma p\to\eta p$
cross section, they have to insert it when explaining the the data
for $\gamma p\to \eta' p$. Finally, the third $S_{11}$
resonance is predicted also in the hypercentric constituent quark model
\cite{GiSa01} which gives $W_r = 1861$ MeV.
In conclusion, the resonance $S_{11}$ which is needed for the description of
the available data \cite{Dug02}
around $K_0 = 1350$ MeV has properties which agree
quite well with the results obtained in the recent theoretical and experimental
investigations.

\section{Perspectives of the model}

Using our model with twelve resonances (together with the third $S_{11}$
resonance), we have extended the calculation of the cross section for $\gamma
p\to\eta p$ to higher photon energies. There are other theoretical predictions
in the energy region above $K_0 = 2$ GeV, from which we would like to note the
Ref.\,\cite{ChiNT02} where the usual isobar model approach was supplemented by
the Regge model. However, the results provided by this hybrid model in the
region $K_0 = 2 - 4$ GeV are not in good agreement with the data. Extending
our model to the energy $K_0 = 4$ GeV with background parameters from
Ref.\,\cite{ChiNP02} we are faced again with unacceptable increase of
the cross section above the energy $K_0 = 3.4$ GeV caused by anomalously large
contribution of the vector meson exchange. Therefore, we have introduced the
new set of $VNN$ parameters (fourth set in Table~\ref{tab2})
and have obtained the result shown by the solid curve in Fig.\ref{fig8}.
Above the photon energy $K_0 = 1.5$ GeV a peripheral $\eta N$ interaction
leading to the excitation of higher resonances $F_{17}(1990)$, $G_{17}(1990)$,
$G_{19}(2250)$, and $H_{19}(2200)$ becomes important. The latter are not well
studied as yet, in particular since relatively large widths of these
resonances (see Table~\ref{tab1})
prevent an isolation in their individual contributions to the
observables. New measurements in the high energy region ($K_0 > 2$ GeV) which
could solve this problem as well as establish the properties of the fourth
$S_{11}$ resonance \cite{CheKa02} are very desirable.

 I am grateful to A.I. Fix.


\newpage
\begin{table}
\renewcommand{\arraystretch}{1.7}
\caption{Resonance parameters obtained by fitting the $\gamma p\to \eta p$
observations to the available data. The background was taken into account as
in Ref.~\protect\cite{ChiNT02}. The values $\xi_\lambda=\sqrt{\frac{k m \Gamma_{\eta p}}
{q W_r}}A_\lambda$ were introduced in Ref.~\protect\cite{BeMu95}.
} \vspace{1cm}
\begin{center}
\begin{tabular}{|c|c|c|c|c|c|c|}
\hline $N^*$ resonance & $W_r$ [MeV] & $\Gamma_r$ [MeV] & $\gamma^E$ [MeV] &
$\gamma^M$ [MeV]&$\xi_{\frac{1}{2}}[$10$^{-1}GeV^{-1}$]& $\xi_{\frac{3}{2}}$
[10$^{-1}GeV^{-1}$]\\
\hline
$S_{11}(1535)$ & 1535 & 158 & 2.16 & --&2.45&--  \\
$S_{11}(1650)$ & 1645 & 140 & -0.620 & -- &0.795&-- \\
$*S_{11}(1825)$ & 1825 & 160 & 0.275 & --&0.308&--  \\
$P_{11}(1440)$ & 1440 & 350 & -- & 0.250&--&--  \\
$P_{13}(1720)$ & 1722 & 145 & -0.105 & 0.430&-0.462&0.642\\
$D_{13}(1520)$ & 1520 & 120 & 0.200 & 0.330&-0.044&0.117\\
$D_{15}(1675)$ & 1673 & 150 & 0.115 & 0.260&0.261&-0.156\\
$F_{15}(1680)$ & 1680 & 130 & 0.045 & 0.050&-0.0003&0.039\\
$F_{17}(1990)$ & 1990 & 375 & -0.056 & -0.4130&0.153&0.150\\
$G_{17}(2190)$ & 2190 & 400 & -0.175 & -0.240&-0.018&0.126\\
$G_{19}(2250)$ & 2250 & 450 & -0.322 & -0.600&0.253&0.105\\
$H_{19}(2220)$ & 2220 & 450 & -0.120 & -0.700 &-0.078&0.111\\
\hline
\end{tabular}
\label{tab1}
\end{center}
\end{table}
\begin{table}[H]
\renewcommand{\arraystretch}{1.7}
\caption{ Vector meson coupling constants used in the present work for the
$t$-channel part of the $\gamma p\to \eta p$ process. }
 \vspace{1cm}
\begin{center}
\begin{tabular}{|c|c|c|c|c|c|c|}
\hline
Meson & Mass [MeV] & $\frac{(g_{VNN}^v)^2}{4\pi}$ &
$\frac{(g_{VNN}^t)^2}{4\pi}$ & $\lambda_V$ & $\Lambda_V$ [GeV] & Ref. \\
\hline
$\rho$ & 770 & 0.50 & 18.6 & 0.89 & 1.80 & \protect\cite{TiBeKa94} \\
$\omega$ & 782 & 23.0 & 0 & 0.192 & 1.40 &  \\
\hline
$\rho$ & 770 & 0.55 & 20.5 & 1.06 & 1.089 & \protect\cite{BeMu95} \\
$\omega$ & 782 & 8.11 & 0.20 & 0.31 & 1.106 &  \\
\hline
$\rho$ & 768.5 & 0.458 & 17.5 & 0.81 & 1.3 & \protect\cite{ChiNP02} \\
$\omega$ & 782.6 & 20.37 & 0 & 0.291 & 1.3 &  \\
\hline
$\rho$ & 768.5 & 0.458 & 1.09 & 0.81 & 1.0 & \protect\cite{ChiNT02} \\
$\omega$ & 782.6 & 6.45 & 0 & 0.29 & 1.3 &  \\
\hline
\end{tabular}
\label{tab2}
\end{center}
\end{table}
 \newpage
\vspace{1cm}

\vspace{1cm}

\vspace{1cm}


\vspace{1cm}

\begin{figure}
\begin{center}
\includegraphics[width=10cm,keepaspectratio]{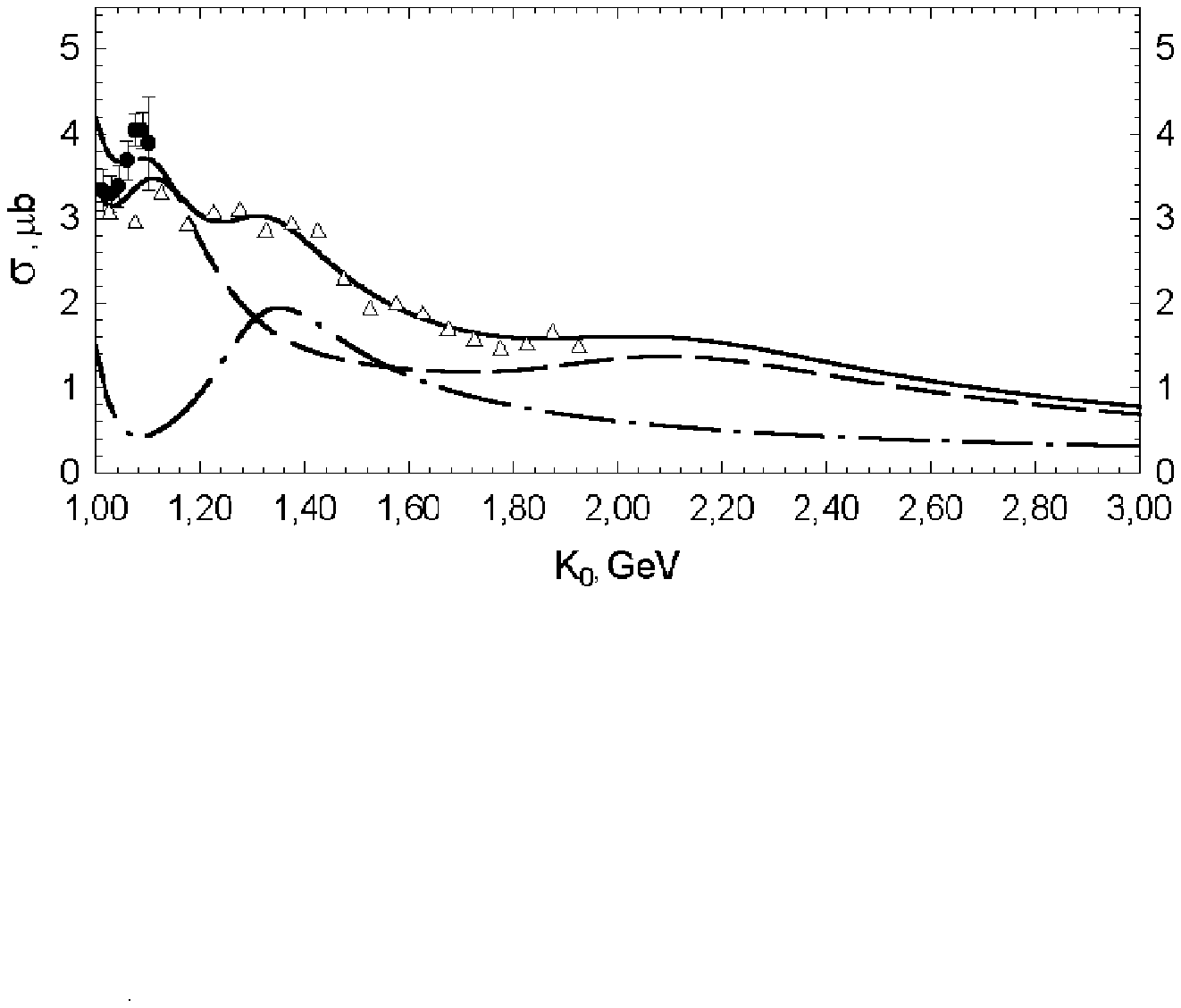}

\vspace{.5cm} \caption{ Integrated cross section for $\gamma p\to \eta p$. The
dashed and solid curves are obtained using the model with eleven and twelve
resonances, respectively, plus background. The dash-dotted curve includes only
$s$-wave resonances from Table~\protect\ref{tab1} plus background. The data are
taken from \protect\cite{Ren02} (circles) and \protect\cite{Dug02} (triangles). }
\label{fig5}
\end{center}
\end{figure}
\vspace{1cm}

\begin{figure}
\begin{center}
\includegraphics[width=10cm,keepaspectratio]{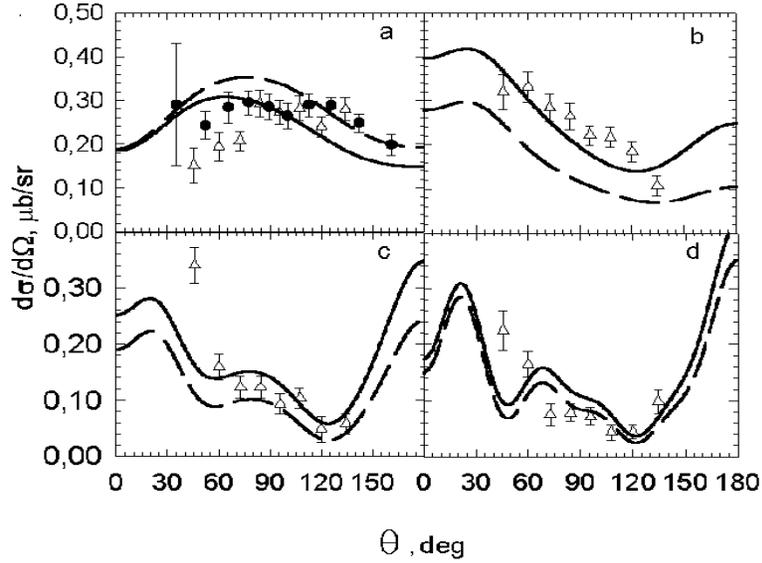}

\vspace{.5cm} \caption{Differential cross section $d\sigma/d\Omega$ for $\gamma
p\to \eta p$ calculated in the c.m. system for the lab photon energies $K_0 = 1025$ MeV (a),
1325 MeV (b), 1625 MeV (c), and  1925 MeV (d). The dashed and solid curves are
obtained within the models including eleven and twelve resonances, respectively. The
data are from \protect\cite{Ren02} (circles) and \protect\cite{Dug02} (triangles).}
\label{fig6}
\end{center}
\end{figure}
\vspace{1cm}
\begin{figure}
\begin{center}
\includegraphics[width=7cm,keepaspectratio]{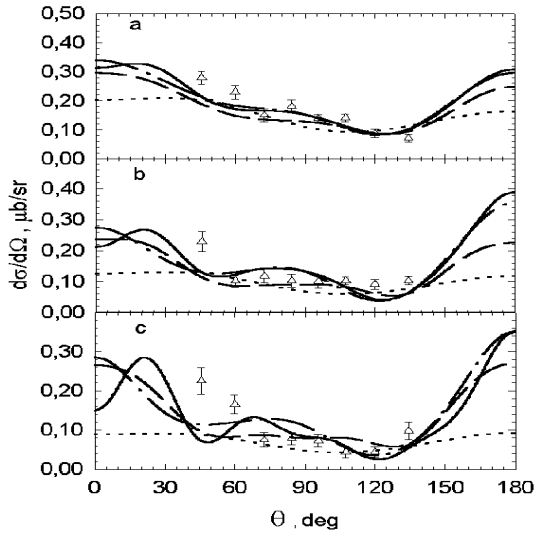}

\vspace{.5cm} \caption{ Differential cross section $d\sigma/d\Omega$ for $\gamma
p\to \eta p$ calculated in the c.m. system for the lab photon energies $K_0 = 1525$ MeV
(a), 1725 MeV (b), 1925 MeV (c). The solid curve is obtained within the models
including twelve resonances. The dashed and dash-dotted curves are obtained within
the models including ten resonances without $F_{17}(1990)$, $H_{19}(2220)$ and
without $G_{17}(2190)$, $H_{19}(2220)$ respectively. The dotted curve is obtained
within the models including eight resonances
(without $F_{17}(1990)$, $G_{17}(2190)$, $G_{19}(2250)$, $H_{19}(2220)$). Data are from
\protect\cite{Dug02}. } \label{fig7}
\end{center}
\end{figure}
\vspace{1cm}


\begin{figure}
\begin{center}
\includegraphics[width=15cm,keepaspectratio]{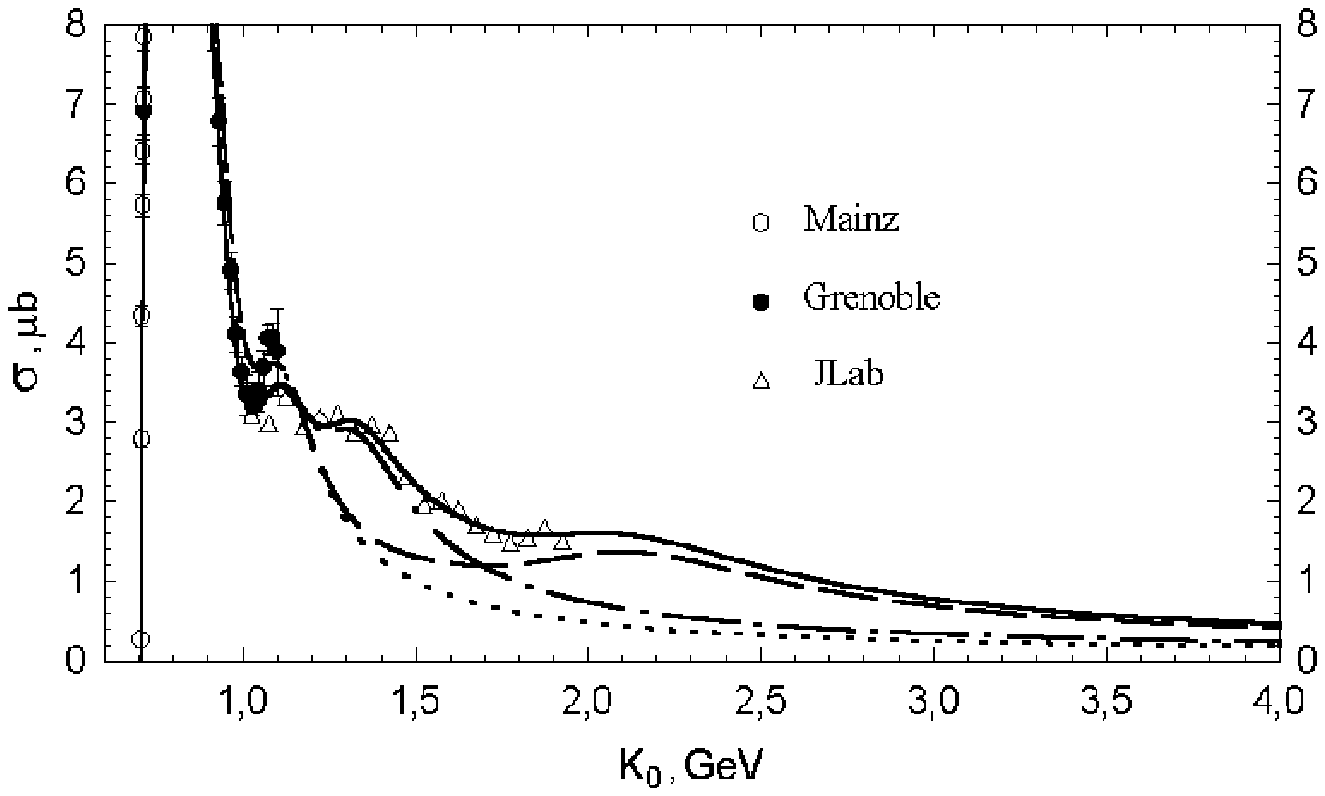}

\vspace{.5cm} \caption{ Integrated cross section for $\gamma p\to \eta p$ as a function
of the lab photon energy$K_0$. The calculation was performed using the model with twelve
resonances (solid curve), with eleven resonances (without $S_{11}(1825)$) (dashed
curve), with eight resonances (without $F_{17}(1990)$, $G_{17}(2190)$,
$G_{19}(2250)$),$H_{19}(2220)$ (dash-dotted curve), and with seven resonances
(without $S_{11}(1825), F_{17}(1990)$, $G_{17}(2190)$, $G_{19}(2250)$), $H_{19}(2220)$
(dotted curve). The background was treated as in the "hybrid" model of
Ref.\,\protect\cite{ChiNT02}. The data are from\protect\cite{Kru95,Ren02,Dug02}. }
\label{fig8}
\end{center}
\end{figure}
\vspace{1cm}


\begin{thebibliography}{99}

\bibitem{CGLN}
G.F. Chew, M.L. Goldberger, F.E. Low and Y. Nambu, Phys. Rev. {\bf 106}, 1345 (1957)

\bibitem{BlWei52}
J.M. Blatt and V. Weisskopf, {\it Theoretical Nuclear Physics}
(N.Y., Wiley $\&$ Sons, 1952)

\bibitem{BeTa91}
C. Bennhold and H. Tanabe, Nucl. Phys. A{\bf 530}, 62 (1991)

\bibitem{BeMu91}
M. Benmerrouche and N.C. Mukhopadhyay, Phys. Rev. Lett. {\bf 67}, 1070 (1991)

\bibitem{TiBeKa94}
L. Tiator, C. Bennhold, and S.S. Kamalov, Nucl. Phys. A{\bf 580}, 455 (1994)

\bibitem{BeMu95}
M. Benmerrouche, N.C. Mukhopadhyay, and J.F. Zhang,
Phys. Rev. D{\bf 51}, 3237 (1995)

\bibitem{FiAr97}
A. Fix and H. Arenh\"ovel, Nucl. Phys. A {\bf 620}, 457 (1997)

\bibitem{HiDe73}
H.R. Hicks, S.E. Deans, D.T. Jacobs, and D.L.Montgomery, Phys. Rev. D{\bf 7}, 2614 (1973)

\bibitem{ChiNP02}
W.-T. Chiang, S.N. Yang, L. Tiator, and D. Drechsel, Nucl. Phys. A{\bf 700}, 429
(2002)

\bibitem{PDG00}
Particle Data Group, Eur. Phys. J. C{\bf 15}, 1 (2000)

\bibitem{TRS01}
V.A. Tryasuchev, Russ. Phys. J. {\bf 44}, 677 (2001);
Phys. Atom. Nucl. {\bf 65}, 1673 (2002)

\bibitem{TRS03}
V.A. Tryasuchev, Russ. Phys. J. {\bf 46}, 403 (2003)

\bibitem{S11}
At this stage the third resonance $S_{11}(1825)$ marked by the asterisk in
Table~\protect\ref{tab1} was not included into the fitting procedure

\bibitem{Kru95}
B. Krusche, J. Ahrens, G. Anton, et al., Phys. Rev. Lett. {\bf 74}, 3736 (1995)

\bibitem{Ren02}
F. Renard, M. Anghinolfi, O. Bartalini, et al., Phys. Lett. B{\bf 528}, 215 (2002)

\bibitem{Ajak98}
J. Ajaka, M. Anghinolfi, V. Bellini, et al., Phys. Rev. Lett. {\bf 81}, 1797 (1998)

\bibitem{Bock98}
A. Bock, et al., Phys. Rev. Lett. {\bf 81}, 534 (1998)

\bibitem{ABBHHM}
ABBHHM Collaboration, Phys. Rev. {\bf 175}, 1669 (1968)

\bibitem{Var98}
G.A. Vartapetyan and S.E. Piliposyan, Sov. J. Nucl. Phys. {\bf 32}, 804 (1980)

\bibitem{Heu70}
C.A. Heusch, C.Y. Prescott, L.S. Rochester, et al., Phys. Rev. Lett. {\bf 25}, 1381
(1970)

\bibitem{ChiNT02}
W.-T. Chiang, S.N. Yang, L. Tiator, M. Vanderhaeghen,
and D. Drechsel, Phys.Rev.C{\bf 68}, 045202 (2003)

\bibitem{Dug02}
M. Dugger, B.G. Ritchie, J. Ball, et al., Phys. Rev. Lett. {\bf 89}, 222002
(2002)

\bibitem{Bai01}
J.Z. Bai, et al., Phys. Lett. B{\bf 510}, 75 (2001)

\bibitem{Sagh02}
B. Saghai, Z. Li, nucl-th/0202007

\bibitem{CheKa02}
G.-Y. Chen, S.S. Kamalov, S.Y. Yang, et al., Nucl. Phys.A{\bf 723}, 447 (2003)

\bibitem{GiSa01}
M.M. Giannini, E. Santopinto, A. Vassalo, Eur.Phys.J.A{\bf 12} 447 (2001)

\end{thebibliography}
\end{document}